UDC 519.83

**Andrzej Jarynowski**

WHICH ACTIVATION FUNCTION OF COOPERATION DESCRIBES HUMAN BEHAVIOR?

*Properties of cooperation's probability function in Prisoner`s Dilemma have impact on evolution of game. Basic model defines that probability of cooperation depends linearly, both on the player's altruism and the co-player's reputation. I propose modification of activation function to smooth one (hyperbolic tangent with scaling parameter a, which corresponds to its shape) and observe three phases for different range of a. (1) For small a, strategies seem to randomly change in time and situation of mixed choices (one cooperates and second defects) dominate. (2) For medium a, players choose only one strategy for given period of time (the common state can switch to opposite one with some probability). (3) For large a, mixed strategy (once defect, once cooperate) is coexisting with common strategies and no change is allowed. I believe that proposed function characterizes better socio-economical phenomena and especially phases 1 and 2 contain most of human behavior.*

**Introduction**.

Optimization problems are well studied in computer science and mathematics and individual human decision can be understood this way [1]. Various methods have been proposed to forecast individual decision including Markov chain models, machine learning, neural networks, Bayesian networks, cellular automaton, but mainly game theory. In a world of social studies modeling in genaral is getting more and more popular. Ability to find patterns in sequences of human decisions is an important component of Artificial Intelligence. Laboratory studies, society observations and computer simulations show that successful pattern-recognition is limited by bounded rationality (so uncertainty must be included in the model). The classical Game Theory model describes optimal strategy known as stable states or Nash equilibrium of cooperation and defect in a single game. Unfortunately, if game is repeated and players have memory and can adopt their strategy, the optimal strategy is to defect. This is happening with loss to society. However sociological studies provide much more variety in human behavior. In order to imitate society, a simple, no-parameter model of the Evolutionary Prisoner Dilemma was previously proposed [2] and developend in [3]. However, limiting only to winning strategies (as vet for vet with vartiation allowing to forgive) in sense of Axelrod Tournament [4], do not reflect all of the observed real situations. Human societies, are organized around altruistic, cooperative interactions [5], while the same time achieving a cooperative solution is very difficult if there is a change for exploitation others players or state [6].

Cooperate or defect: these questions can be answered according to some mathematical rules. The setting is described by players acquiring reputation and altruism, which in turn determine their choice of strategy. The probability of cooperation depends, both on the player's altruism and the co-player's reputation. Agents can establish the best strategy in repeated games. Each time a player cooperates, his reputation goes up (vice versa in case of defection). Two key factors [7] are named in human decision making process (estimates of how important each of those factors are very subjective): the normative (Homo Sociologicus) and the rational one (Homo Economicus). I focus on sociological (normative) perspective, there collective behavior is observed. This paper proposes a modification of the activation function used for defining the probability of cooperation in Prisoner's Dilemma (game theory). By changing the parameter $a$ of the activation function, different behaviours can be simulated.

Each agent $i$ is endowed with two parameters: altruism $\varepsilon_i$ and reputation $W_i$. Initial values of the parameters are selected randomly from homogeneous distributions: $\rho(\varepsilon_i)$ is unitary for $-0.5<\varepsilon_i<0.5$, otherwise $\rho(\varepsilon_i)=0$, and $\rho(W_i)$ is unitary for $0<W_i<1$, otherwise $\rho(W_i)=0$.

Each time a player cooperates, his reputation goes up (vice versa in case of defection) but altruism is constant. If agents play in pairs, they choose both only one strategy in almost all cases. The probability that $i$ cooperates with $j$ is given by $P(i,j) = F(\varepsilon_i + W_j) = F(x)$. Reputation change dynamic rules are defined by: if $j$ cooperated, her/his reputation transformed as $W_i \rightarrow (1+W_i)/2$, otherwise $W_i \rightarrow W_i/2$.

Various activation functions could be used:
  - standard case $F(x) = 0$ if $x<0$, $F(x) = x$ if $0<x<1$ and $F(x) = 1$ if $x>1$
  - normalized to probability case $F(x) = (\varepsilon_i + W_j + 0.5)/2$
  - our smooth function $F(x,a) = (1 + \tanh(a(\varepsilon_i + W_j - 1/2)))/2$, where $a$ is scaling parameter.

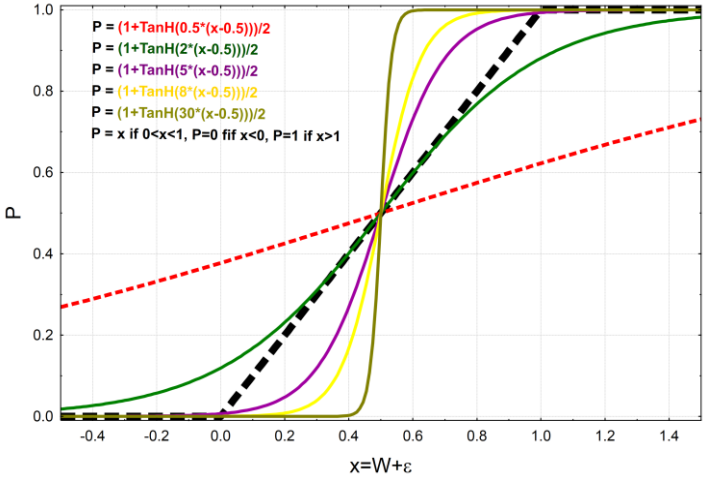

**Fig. 1.** Visualization of different variants of cooperation probability functions in function of sumarized players altruism and co-player reputation.

**Motivation and Implementation.**

Properties of cooperation's probability function in Prisoner Dilemma have impact on evolution of game. Authors of [2] assumed, that $F$ was a linear function of both: the player's altruism in range (-0.5, 0.5) and the co-player's reputation in range (0, 1). Accordingly, the range of values of in range $F$ within (-0.5, 1.5) was limited to (0,1) as follows: the result above 1 was set to 1 and the result below 0 was set to 0. If agents play in pairs, both could choose both dominating strategy in almost all cases [3, 8]. Here, I propose a modification of this probability function to a smooth one (hyperbolic tangent with scaling parameter $a$, which corresponds to shape of curve) and observe three phases for different range of $a$ [Fig. 1].

Here, for neutral altruism (the altruism - $\varepsilon$ of all agents is set to zero) the probability $P(i,j)$ that agent $i$ cooperates with agent $j$ is assumed as: $F(x,a) = (1 + \tanh(a(W_j - 1/2)))/2$ where $0 < W_j < 1$ is the reputation of agent $i$ in eyes of $k$ and altruism is neutral. Parameter $1/a$ can be understand as a human noise [9]. The main observable of our dynamica system is mean reputation of our pair of players $W = (W_j + W_i)/2$. I examime its time evolution for a single pair and collective statistcs for all possible configuration of initial conditions.

Note that in the limit of infinite $a$, $P$ is stepwised and game results are are fully determinated by initial contidtions. The most important scenario shows up for initial condition $W\sim1/2$, because the is no common strategy between palyers. I observe three phases for different range of $a$, which could explain different non-deterministic social behaviors [Fig. 2]. Boundaries between phases are smooth and some properties overlap.

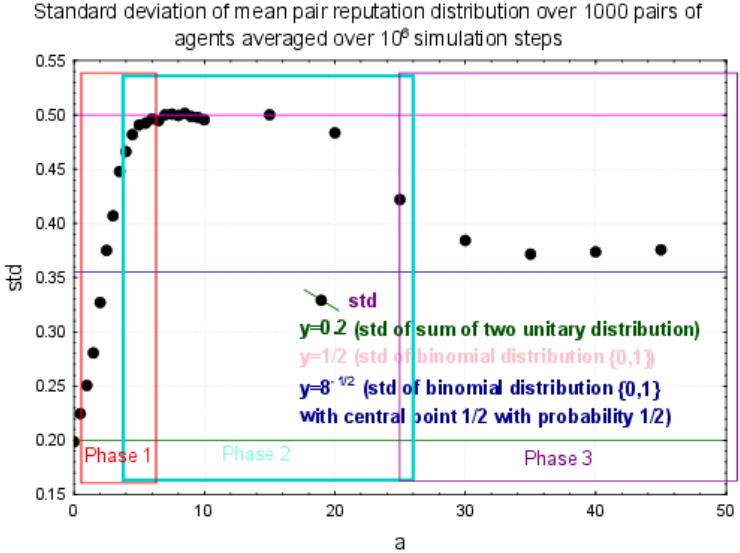

**Fig. 2.** STD of reputation in population with characteristic lines for comparable distributions and 3 distingished phases

Discussed competing strategies (called 'strange strategy') for $W\sim1/2$ for finite $a$ are not stable and their lifetime is also invetsigated [Fig. 2, 3]. For infinite $a$, the game has three possible outcomes: a) both cooperate (probability 0.25), b) both defect (probability 0.25) and c) a cyclic series of games where either $j$ cooperates and $i$ defects, or the opposite, exchanging the strategies at each time step – 'strange strategy'. Then, the distribution of $W$ from initial conditions consists of three pheses at $\{0, ½, 1\}$ [Fig. 3]. Small perturbations within the phase boarder will self-correct back to these fixed points. Concluding, these fixed points are stable and are also attractors of this system. However, its is not as simply in finite $a$ case. In this paper, I rised some questions as: What is the resulting equilibrium state? How many regimes $a$ form? What is the composition of each phase in various regimes?

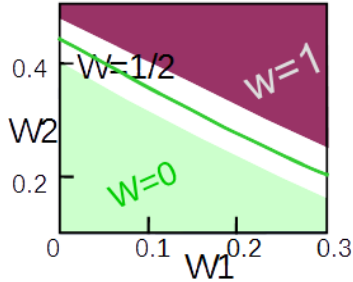

**Fig. 3** Part of phase diagram for deterministic (very large) range of *a* in function of initial condition (reputation of both players). We observe 3 pure phases W=1 (always cooperate), W=0 (always defect) and in the middle W=1/2 (once defect once cooperate – 'strange strategy').

**Results – Phases.**

I examine the model of cooperation in the Prisoner's Dilemma with a new (smooth) activation function containing a scaling parameter *a*. As the result I distinguished three scenarios of behavior depending on different values of *a*. Starting with with $a = 0$ (activation function is constant and does not depent on any variables), first weak correlated very noisy regime (1) is reached. As regime (1) crossover (2) rather rapidly around $a = 5$ with sharp bound. Regime (2) goes into (3) very slowly and transition is somewhere above $a = 25$, where system become deterministic.

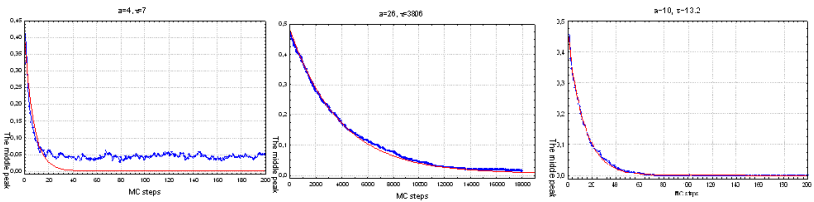

**Fig. 4.** Fits of decay function with intensity of the middle peak for selected characteristic *a*

There are three modes of behavior (termed phases for brevity from now on) for different ranges of *a*. However, the boundaries between these phases are fuzzy: some properties overlap.

*(Regime 1)* - For $a < 5$, strategies seem to randomly changing in time like generalized mean-reverting quasi-geometric Brownian motion with atractive boundaries [Fig. 5]. Apart from a noisy base, atraction to mean and boundaries is noticed. However, autocorrelation (memory) is growing with *a*.

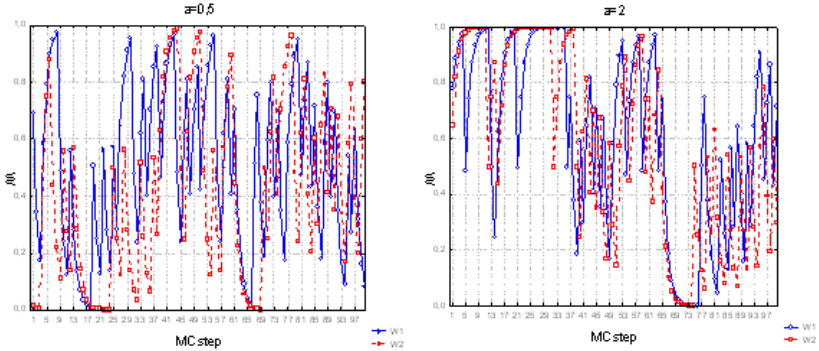

**Fig. 5.** Regime 1. For small a reputations evolve quite randomly

*(Crossover Regime 1&2)* - For *a* around 5, players choose in most cases the common strategy (both cooperate or both defect) and play this for some time. A state can switch to the opposite one with some probability, as shown in [Fig. 6].

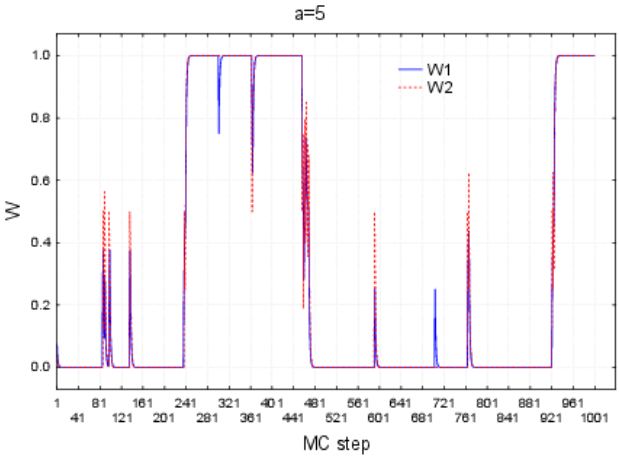

**Fig. 6.** Crossover Regime 1&2. Characteristic depolarisation

*(Regime 2)* - For $5 < a < 25$ the common strategies dominate. During a few initial steps of the simulation usually the players quickly choose some common strategy. Still, some 'strange' strategy is also possible, where mean reputation of both players is around 0.5, i.e. close to the center of the range [Fig. 8 left]. This means that when one of players cooperates, the other defects; in the next step the roles are exchanged, and so on. Yet after some time, only the common strategies survive. The disappearance of the 'strange' peak of mean reputation can be described as exponential decay $\exp(-t/\tau)$ [Fig. 4, 7, 9]. The strange oscillating scenario cannot persist because the system is not fully deterministic. Namely, it is always possible that the cycle is broken by an error: an agent selects a strategy despite its small probability. In a consequence, one of two common strategies prevails. The best fit of the exponential decrease of the strange behavior is around $a = 10$ [Fig. 9]. In general, the intensity of the 'strange' peak does not decrease to zero for smaller $a$, because the probability of switching back to the strange state remains positive. On the other hand for larger $a$ the relaxation time is very large.

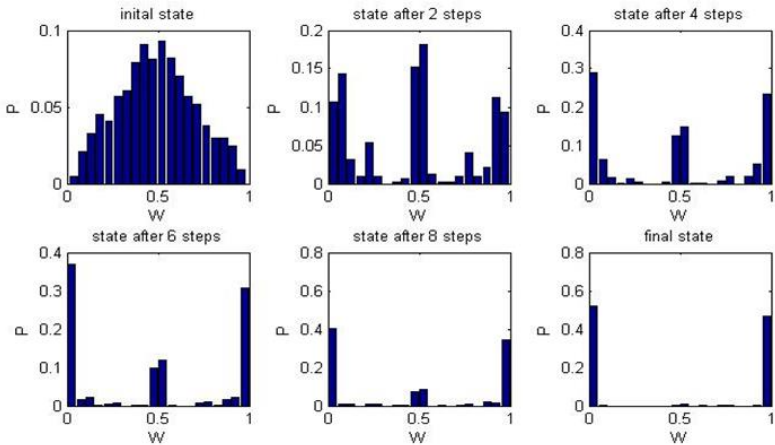

**Fig. 7.** Histograms of mean reputation of 1000 pairs in consecutive time steps for a=8

*(Regime 3)* - For $a > 25$ the spectrum of strategies does not vary in time. I observe pairs of agents who play the common strategy: both cooperate or both defect. This happens for a half of the simulated population (0.25+0.25). For the remaining half of population, the mean reputation is 0.5 [Fig. 2], what reflects the oscillating of strategies. The probability of this 'strange' strategy does not decrease in time and it coexists with the common strategies [Fig. 8 right]. Asymptotically, for infinite $a$, the probability function $P$ turns into the stepwise one and the system is no stochastic any more. In this situation, the time evolution can be predicted from the initial state. In particular, the 'strange' strategy is a consequence of the initial state where one player has reputation above 0.5, and the other below 0.5. In each step, one player loses his reputation but another gains, in the next step the opposite and so on.

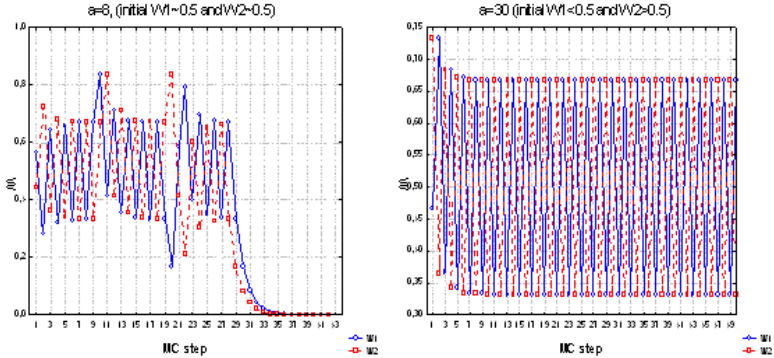

**Fig. 8.** Regime 2 (left), Phase 3 (right). Cycles around mean W=0.5

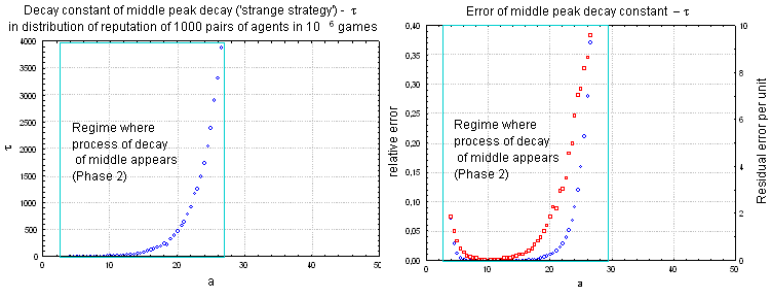

**Fig. 9.** Estimation of decay constant $\tau$ representing 'strange' strategy - phase 2 (left), estimation error of $\tau$ which has minimum around $a$=10 (right)

### Conclusions & Speculations.

A simple model from game theory, which can imitate a decision making patterns, is proposed. I explore the model of cooperation in the Prisoner's Dilemma, based on reputation [2]. Investigated probability $P(i, j)$ that agent $i$ cooperates with agent $j$ is assumed as $P(i, j, a) = \{1 + \tanh[a(W_j - 1/2)]\}/2$, where $1/a$ is a measure of errors of the players (uncertainty of the game). In the limit of infinite a, the game is deterministic with possible outcomes: a) both cooperate (W=1), b) both defect (W=0) and c) and strange strategy' (W=1/2). For finite values of *a*, the probability of c) decreases exponentially in time. For small values of *a*, a crossover is observed from the state where only options a) and b) appear (*a*=5) to a homogeneous distribution of W at the most fuzzy case *a*=0.

In this paper, only neutral altruism case was analysed, but other possible values of $\varepsilon$ are just linear shift in the argument of function $F(x)$. The system is equivalent to pair of nodes with a single link between agents. Such a simplest possible interactive system is necessary to understand basic properties of activation function. Others configuration: triangle and networks (fully connected graph, lattices, E-R, B-A, small world or real social networks) were tested for a base model [8]. In literature smooth function of opinion is also described as a Fermi function [10, 11], but I choose tanh due to known and simple mathematical properties. $1/a$ understood as an 'human error' is responsible for spontaneous changes of individual decision and lead to synchronized change of global strategy [9]. It reflects the dynamicity of real social system better than standard equilibrium approach. Proposed function characterizes different phases, which can be applied to social phenomena:

- In phase 1 process of decision making is very sensitive to condition and people are not consistent in their strategies. No consensus, but also no conflicts are possible in long terms (e.g. children games [12]).

- In crossover 1&2 phases people act in schisophemic way (once are very consequent in one strategy, to change it rapidly to second one (e.g. Dr. Jekyll and Mr. Hyde).

- In phase 2 one strategy dominate for long time, but very important issue could change it (e.g. Nazism in Germany [13]).

Concluding, dominat startegy in game can change over time according to a set of fixed rules and presente parameter *a* determines how one state of the system moves to another state.

*Moldova State University, Kishinev, R. of Moldova*

*Smoluchowski Institute, Jagiellonian University, Cracow, Poland*

*E-mail:* ajarynowski@gmail.com

IN RUSSIAN

**Ярыновский, А. Молдавский Государственный Университет в Кишиневе**
КАКАЯ ФУНКЦИЯ АКТИВАЦИИ СОТРУДНЕЧЕСТВА ОПИСЫВАЕТ ЧЕЛОВЕЧЕСКОЕ ПОВЕДЕНИЕ?

*Свойства функции вероятности сотрудничества в дилемме заключённого оказывают влияние на эволюцию игры. Базовая модель определяет, что вероятность сотрудничества линейно зависит как от альтруизма игрока так и от репутации остальных игроков. Предлагаю изменenить функции активации на гладкую функцию (гиперболический тангенс с параметром масштабирования a, который соответствует его форме) и наблюдать три*

фазы различного диапазона a:

(1) Для малых a, стратегии случайно изменяются во времени и ситуация смешанных вариантов (один сотрудничества и второй дефект) преобладает;
(2) Для средних a, игроки выбирают только одну стратегию для определенного периода времени (общее положение может перейти на противоположную с некоторой вероятностью);
(3) Для больших a, смешанная стратегия (один раз дефект, один раз сотрудничество) сосуществует с общими стратегиями и изменения не допускаются.

*Я считаю, что предложенная функция лучше характеризует социально-экономические явления и особенно фаза 1 и 2 включают в себя большую часть поведения человека.*
***Ключевые слова: теория игр, вычислительные социальные науки***

*IN UKRAINIAN*
**Яриновскі, А. Молдавський Державний Університет в Кишиневі**
Яку функцію активації співробітництва описує поведінку людини?
***Ключові слова: теорія ігор, обчислювальні соціальні науки***